\documentclass[aps,apl,groupedaddress,twocolumn]{revtex4}

\usepackage{graphicx}
\begin{document}

\title{Free-space quantum key distribution with entangled photons}

\author{Ivan Marcikic}
\author{Ant\'{\i}a Lamas-Linares}
\author{Christian Kurtsiefer}


\affiliation{Department of Physics, National University of
Singapore,
  Singapore, 117542}

\date{\today}

\begin{abstract}
We report on a complete experimental implementation of a quantum key
distribution protocol through a free space link using
polarization-entangled photon pairs from a compact parametric
down-conversion source. Over 10 hours of uninterrupted communication
between sites 1.5~km apart, we observe average key generation rates
of 630 per second after error correction and privacy amplification.
Our scheme requires no specific hardware channel for synchronization
apart from a classical wireless link, and no explicit random number
generator.
\end{abstract}


\maketitle

Quantum key distribution \cite{gisin:02} is probably the most mature
application developed out of quantum information science during the
last decade. Based on initial ideas of Wiesner \cite{wiesner:83},
quantum states of physical systems with small Hilbert spaces can be
used to encode information in a way that an illegitimate attempt of
accessing that information will result in a perturbation of the
state, and consequently in revealing the interception attempt. Thus,
the secrecy of a bit string or a key between two parties could be
ensured relying on fundamental laws of quantum mechanics rather than
assumptions on the mathematical complexity of problems like
factoring.

 A specific protocol to establish a secret key between two
parties was established by Bennett and Brassard (BB84)
\cite{bennett:84}, which used the polarization degree of freedom of
single photons. An alternative scheme for a quantum optics based key
distribution was suggested by Ekert \cite{ekert:91}, where nonlocal
correlations in the measurements on entangled photon pairs both
allow to establish a secret key and evaluate the knowledge of an
eavesdropper out of a violation of Bell inequalities. In this
scenario, only the detection units needed to be in possession of
legitimate communication partners, while the entangled photon source
need not to be in a trusted hand. Work on security analysis of
quantum key distribution systems also typically makes use of
entanglement, even in the case of the original or modified BB84
implementations with faint coherent pulses \cite{theorysec}.

Practical implementations of quantum key distribution (QKD) can be
classified according to the type of protocol and the physical
transmission channel. We distinguish protocols of the prepare and
send (PaS) type and entanglement based schemes. The transmission
channel is either optical fiber or free space.

Most demonstrations of QKD (including all commercial systems) are
PaS protocols implemented over fiber channels \cite{PaSfiber} and a
few over free space \cite{PaSfreespace}. The choice of channel
reflects the maturity of fiber technology and its commercial
possibilities in current networks. The preference for PaS protocols
is due to the technological simplicity compared with entanglement
based schemes. However, security in PaS protocols relies on the
availability of high bandwidth trusted random numbers, which does
not need to be the case for entanglement based systems.

First steps towards entanglement based QKD were experiments on
distributing entanglement in the field over fiber links
\cite{tittel:98} and in free space \cite{resch:05,peng:05}. More
recently, QKD over a purposely laid fiber link was reported
\cite{poppe:04}. In this paper we describe a entanglement based
(modified BB84) full field implementation of QKD over an ad hoc free
space link. The system is based on a compact ($80\times
50$\,cm$^{2}$) spontaneous parametric down-conversion source (SPDC),
compact detection modules, a free space standard wireless internet
protocol link and a software synchronization protocol taking
advantage of intrinsic time correlations in the SPDC process. Error
correction and privacy amplification are implemented on the fly
producing a continuous stream of secure key.

A schematic experimental set-up is shown in figure~1. Light at
404\,nm emitted by a cw laser diode (LD) is circularized and focused
(PO) to a waist of 90\,$\mu$m into a $\beta$-barium borate (BBO)
non-linear crystal. There, polarization-entangled photon pairs are
created via type-II SPDC out of 50\,mW power. An additional
half-wave plate (WP) and BBO crystals (CC) compensate walk-off
effects~\cite{kwiat:95,trojek:04}. These crystals are also used to
set the relative phase between horizontal (H) and vertical (V)
polarizations such that the source produces photon pairs in a
singlet Bell state:
\begin{eqnarray}
\left| \Psi^- \right\rangle =\frac{1}{\sqrt{2}}(\left|
H\right\rangle _{A}\left| V\right\rangle _{B}-\left| V\right\rangle
_{A}\left|H\right\rangle _{B}) \nonumber
\end{eqnarray}

Indices $A$ and $B$ (denoted Alice and Bob) represent two spatial
modes defined by coupling SPDC light into single mode optical fibers
(SMF)~\cite{trojek:04}. Center wavelength and spectral bandwidth are
805.2\,nm and 6.3\,nm for Alice's mode, and 810.7\,nm and 6.8\,nm
for Bob's mode. Each fiber passes through a polarization controller
(FPC) to undo fiber induced polarization transformations. At the
source we observe a coincidence rate of $24\,000~{\rm s}^{-1}$ with
an overall coupling and detection efficiency of 22\,\%, detected by
passively quenched Silicon avalanche photo diodes (Si-APD). The
entanglement quality is verified by measuring polarization
correlations in H/V and $+45/-45$ basis. We observe a visibility of
$98\pm2.6$\,\% and $92\pm2.2$\,\%, respectively.

\begin{figure}[h]
\includegraphics {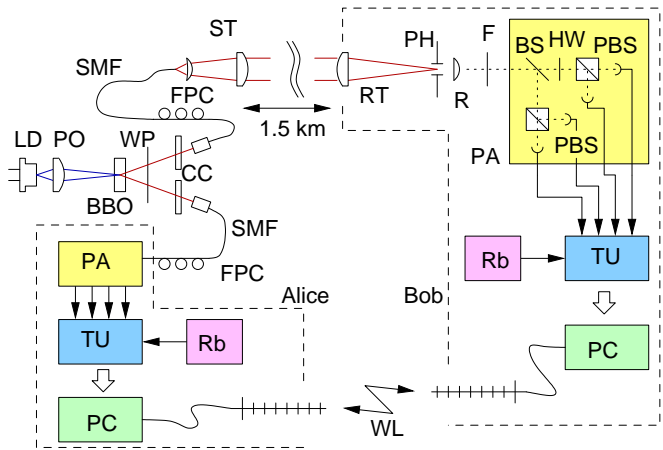}
\caption{{Scheme of the experimental set-up. Polarization-entangled
photons are distributed between two parties, Alice and Bob,
separated by 1.5\,km. Coincidence identification uses time-stamp
units and requires a wireless link.}} \label{setup}
\end{figure}

In a BB84-type QKD experiment, Alice's and Bob's polarization
detection units (PA) randomly analyze the received photons in two
maximally conjugated basis (H/V and +45/-45). For this purpose we
use compact detection units~\cite{kurtsiefer:02} relying on a
non-polarizing beam splitter (BS) to ensure the random choice of the
measurement basis. Projection onto +45/-45 basis in the transmitted
arm is via a half wave plate (HW) and polarizing beam splitter
(PBS); in the reflected arm the H/V projection is implemented
directly with a PBS. Photons are finally detected with four Si-APDs
cooled to around $-15\,^{\circ}$C (with outside T=$28^{\circ}$C)
with an average dark count rate of 1\,000\,s$^{-1}$ per detector.

Optical ports are situated on the rooftops of two buildings in the
campus of National University of Singapore (E103$^\circ$ 46' 48.3'';
N1$^\circ$ 17' 48.7'' and E103$^\circ$ 46' 3.5''; N1$^\circ$ 18'
8''), separated by 1.5\,km. The entangled pair source is connected
by 50\,m of single-mode fiber to the sending telescope (ST)
collimating the fiber mode into a Gaussian beam with a waist of
15\,mm. The receiving telescope (RT, focal length of 310\,mm,
diameter 77\,mm) focuses light onto a pinhole (PH) with 50\,${\rm
\mu}$m diameter acting as a spatial filter. An interference filter
(F, $\lambda_0=810.7$\,nm, FWHM$=5$\,nm) is used for background
suppression. The pinhole is imaged (R) onto the Si-APDs in the
polarization analyzer. The pointing accuracy of the telescopes is
$\approx$10\,${\rm \mu}$rad.

In a protocol based on pairs of photons, it is fundamental to
identify reliably which events are correlated in their times of
arrival. With a cw laser pumped source, pair arrival times are
random on all time scales, both due to the preparation process and
the randomizing influence of losses. In lab experiments and some
field implementations, coincidence or timing information required a
dedicated hardware channel \cite{poppe:04,peng:05}. In our
experiment, we use a software-based coincidence identification,
where we continuously register the detection time of all photoevents
on both sides with a timestamp unit (TU) locked to a Rb oscillator
\cite{jennewein:00,kurtsiefer:02,resch:05}.

Before coincidences can be identified, the clocks on both sides need
to be synchronized to the order of a coincidence time window. We
start with a standard NTP protocol \cite{mills:91} between the
controlling hosts (PC in fig.~1) to an accuracy $<100$~ms, followed
by a tiered cross correlation on raw photodetection timings. Initial
locking of the remote clocks takes a few seconds to extract coarse
and fine timing information with a resolution of 2.048\,${\rm \mu}$s
and 2\,ns, respectively, followed by an FFT algorithm to find the
maxima of the cross correlation functions.

For efficient timing communication, we partition detection events in
packets every $2^{29}$\,ns and encode time intervals between
consecutive events, reaching a bandwidth $\approx$\,13\,\% above the
Shannon limit or 1~Mbit per second for 50\,000 events per second.
Timing information consumes the largest bandwidth of all
communication on the classical channel, but is comparable to schemes
with a fixed timing raster which would not allow to work efficiently
with a cw pumped source. We used a standard 801.11g wireless
connection (WL) for classical communication.

\begin{figure}[h]
\includegraphics [width=8.6cm]{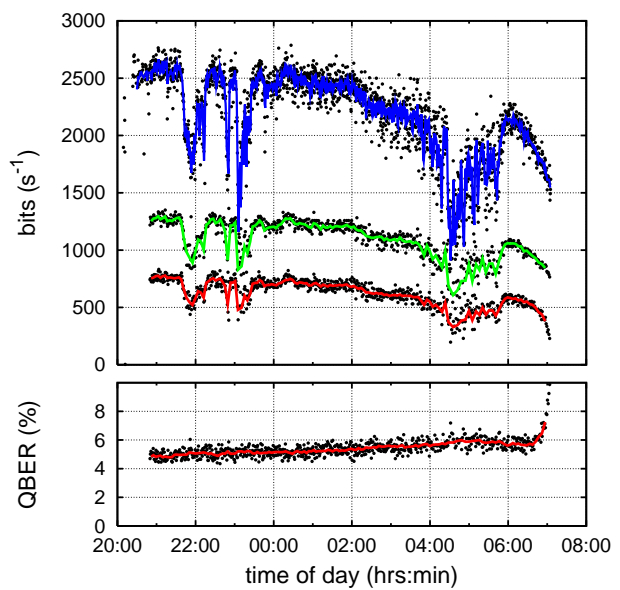} \caption{Quantum key distribution between two sites 1.5\,km apart.
In the top panel, the traces correspond to the raw coincidence rates
between the remote sites, sifted and final secure key after error
correction and privacy amplification. The bottom panel shows the
measured quantum bit error ratio of the sifted key. The acquisition
took place during an uninterrupted period of $\approx$\,10~hours at
night. The reduced scatter in the sifted and error corrected keys is
due to the clustering of the data for error correction and privacy
amplification. For clarity, the traces show a downsampled (1 out of
10) subset.} \label{results}
\end{figure}

The coincidence identification is carried out asynchronously on the
high count rate side. Due to timing jitter of photodetectors,
reference clock noise, clock drift servo noise, and noise in time
stamp units, we observe a cumulative width of 1.4\,ns (FWHM) for the
distribution of coincidence time differences $\Delta t$. For raw key
generation, we accept coincidences with $|\Delta t|<1.75$\,ns.
Coincidences with $|\Delta t|<3.75$\,ns are used to servo the clock
drifts for continuous operation beyond the reference clock
stability, which would only allow for only a few minutes of
operation. Accidental coincidences (8.63\,s$^{-1}$ in average between
21:00 and 06:00 hours) monitored in a 3.75\,ns wide reference window displaced
by 20\,ns from the main coincidence window are consistent with total
single event rates on transmitter side (99\,692\,s$^{-1}$) and receiver side
(18\,325\,s$^{-1}$) detectors. To ensure that different timing delays of
individual detectors could not be exploited for eavesdropping, we
equalized the average delays to $\approx$\,0.25\,ns.

We remove errors in the raw key on the fly with a modified {\sc
CASCADE}/{\sc BICONF} error correction algorithm (following largely
\cite{EC}) on clusters of collated raw key packets with at least
5\,000 bits. The required bandwidth on the classical communication
channel is small compared to the initial sifting.

Privacy amplification \cite{bennett:88} for obtaining $s$ secure key
bits out of $r$ raw key bits in a cluster removes possible knowledge
of an eavesdropper out of an attack (estimated from the observed
error fraction $\eta$ in the raw key), and due to the $c$ revealed
parity bits in the error correction process. We assume an
attack-based knowledge of an eavesdropper of
$e=r/2[(1+z)\log_2(1+z)+(1-z)/\log_2(1-z)]$, with
$z=2\sqrt{\eta(1-\eta)}$. The compression matrix for privacy
amplification to $m=r-e-c$ final key bits is generated on the fly
from a publicized seed for a pseudo random number generator for
every cluster. All observed errors are assigned to an eavesdropping
attempt, no assumptions on the inability of an eavesdropper to
access intrinsic errors in the system are made.

Our results are presented in figure~2. The initial remote
coincidence rate is 2\,600\,s$^{-1}$ and drops to 2\,000\,s$^{-1}$
after $\approx$\,10\,hrs without any intervention or active
stabilization after initial alignment. The secret key exchange is
finally interrupted by the rising sun saturating the detectors. The
quantum bit error ratio (QBER), i.e. the ratio between wrong over
correct events, increases slowly from $\approx$\,5\,\% to
$\approx$\,6\,\%, with an overall average of 5.4\,\%. We do not
observe any appreciable increase in QBER due to the propagation
through the atmosphere. We attribute the small increase of 0.5\%
between the local and remote QBER to residual uncompensated
birefringence in the fiber connecting the source to the telescope.
After the basis reconciliation the averaged sifted and secret keys
bit rates are $1\,100$\,s$^{-1}$ and $630$\,s$^{-1}$. Our error
correction implementation resulted in a residual bit error ratio
(BER) of $2\times 10^{-5}$ in this run, clearly bunching in
particular clusters. After privacy amplification, this lead to 230
clusters with nonidentical final key out of 7\,500.

The remote coincidence rate is around 11\,\% of the local rate. We
observe a transmission between the entangled pair source and the
sending telescope of 85\,\%, through telescope optics 90\,\% and
through interference filter 50\,\%. Separation of 1.5\,km leads to
an additional signal reduction of 50\,\%. Each detection unit
exhibits another 20\,\% of losses. We do not observe a significant
contribution to the losses due to scintillation.

For a second experiment, we replace the interference filter with a
long pass filter (RG780). Over 6 hours of measurement, around 850
bits per second of secret key are distributed on average. The sifted
key bit rate increases to 1\,600\,s$^{-1}$ with an average QBER of
5.75\,\%. The BER before privacy amplification was $2\times 10^{-6}$
on cluster sizes $>10\,000$ bits, leading to 21 nonidentical key out
of 1\,720 after privacy amplification.

In conclusion, we have demonstrated a real-time free-space quantum
key distribution based on a BB84-type protocol with
polarization-entangled photons. A software base coincidence
identification scheme was implemented, not relying on a dedicated
hardware channel. A nighttime experiment with an interference filter ran
uninterruptedly for over 10 hours producing an average secure key rate
of 630 bits per second, while with a long pass filter we observed 850 bits per
second over 6 hours. We believe that this QKD set-up can be used even for
daylight operation: from preliminary experiments we estimate that the
accidental coincidence rate therefore must be reduced by one order of
magnitude, which can be achieved with stronger spectral and spatial filtering,
and a shorter coincidence time window.

The authors acknowledge the DSTA Singapore for their financial
support and thank Darwin Gosal, Gleb Maslennikov, Alexander Ling,
Hou Shun Poh and Meng Khoon Tey for their help.

\end{document}